\documentclass[aps,prl,showpacs,floatfix,twocolumn,amsmath,amssymb,preprintnumbers]{revtex4-1}
\usepackage{mathrsfs}
\usepackage[figuresright]{rotating}
\usepackage{amsmath}
\usepackage{amssymb}
\usepackage{graphicx}
\usepackage{color}
\usepackage{dcolumn}
\usepackage{bm}
\usepackage[breaklinks=true,colorlinks=true,linkcolor=blue,urlcolor=blue,citecolor=blue]{hyperref}

\usepackage{soul}

\makeatletter
\newcommand{\rmnum}[1]{\romannumeral #1}
\newcommand{\Rmnum}[1]{\expandafter\@slowromancap\romannumeral #1@}
\makeatother

\begin{document}

\title{$^{75}$As NMR study of the antiferromagnetic Kondo lattice compound CeNiAsO}

\author{Fangjun Lu$^{1}$}
\author{Xiaobo He$^{1}$}
\author{Kangqiao Cheng$^{1}$}
\author{Zhuo Wang$^{1}$}
\author{Jian Zhang$^{1}$}
\author{Yongkang Luo$^{1}$}
\email[]{mpzslyk@gmail.com}
\address{$^1$Wuhan National High Magnetic Field Center and School of Physics, Huazhong University of Science and Technology, Wuhan 430074, China;}

\date{\today}

\begin{abstract}

We revisit the magnetic properties of the antiferromagnetic Kondo lattice CeNiAsO by $^{75}$As nuclear magnetic resonance measurements. Our results confirm two successive antiferromagnetic transitions of Ce moments at $T_{N1}=9.0(3)$ K and $T_{N2}=7.0(3)$ K. Incommensurate and commensurate antiferromagnetic orders are suggested for $T_{N2}<T<T_{N1}$ and $T<T_{N2}$ respectively, consistent with previous neutron and muon experiments. A Knight shift anomaly, characterized by the failure of $K(T)-\chi(T)$ scaling, is observed below $T^*\sim15$ K, which gives a measure of the onset of coherent $c-f$ correlations. This energy scale is further confirmed by the spin-lattice relaxation rate ($1/T_1$). The analysis of spin dynamics also reveals a quasi-two-dimensional character of spin fluctuations in CeNiAsO. This work paves the way for further $^{75}$As nuclear magnetic resonance studies under pressure.

\end{abstract}


\maketitle

\section{\Rmnum{1}. Introduction}

The ZrCuSiAs-structured rare-earth oxypnictides not only boosted a new wave of research boom on high-$T_c$ superconductors\cite{Hosono-LaOFFeAs,Chen-CeFeAsO,ChenX-Sm1111_F,WangC-Gd1111_Th}, but also provided a novel playground for investigating strongly correlated electronic effects and quantum critical phenomena. Depending on the specific chemical element in the $Tm$ and $Pn$ sites, the family of Ce$TmPn$O ($Tm$=transition metal, $Pn$=P, As) displays a wide spectrum of physical properties. The member CeFeAsO shows a spin-density-wave (SDW) type antiferromagnetic (AFM) transition of Fe-$3d$ electrons near 140 K, and superconductivity can be induced by appropriate F$^-$ doping when the SDW order is suppressed\cite{Chen-CeFeAsO}. At a much lower temperature $\sim 4$ K, the Ce local moments also order antiferromagnetically\cite{Chen-CeFeAsO,ZhaoJ-CeFeAsO_F2008,Cruz2010,Luo2010prb}. In contrast, the analogue compound CeFePO is a non-superconducting heavy-fermion metal with pronounced ferromagnetic correlations\cite{Bruning-CeFePOFMHF,Luo2010prb,Kitagawa-CeFePO,Lausberg-CeFePOAvoid,jesche2017avoided}. For $Tm$=Co, strong interplay between Co-$3d$ and Ce-$4f$ magnetism was observed in both CeCoPO\cite{Krellner-CeCoPO} and CeCoAsO\cite{Sarkar-CeCoAsO}, where the Co-$3d$ electrons show ferromagnetic orders near 75 K, while Ce-$4f$ electrons are on the border to magnetism with enhanced Sommerfeld coefficient. The appearance of $3d$ magnetism in these  compounds complicates the study of the $4f$ physics.

\begin{figure*}[!ht]
\vspace*{-120pt}
\hspace*{-0pt}
\includegraphics[width=18.0cm]{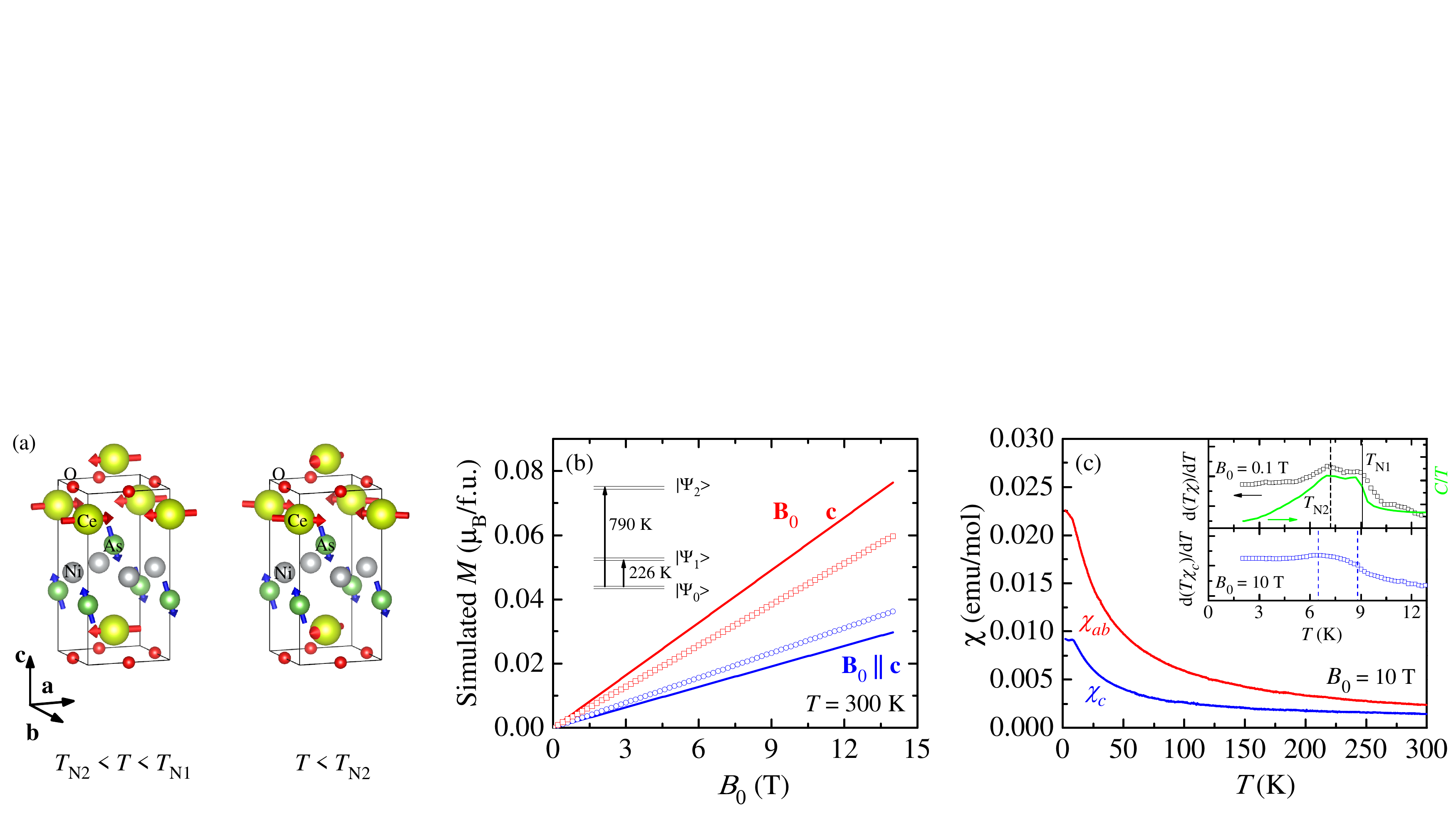}
\vspace*{-20pt}
\caption{(a) Local environment of As in CeNiAsO. The red arrows denote Ce moments, which order in an incommensurate SDW with $\mathbf{q}=(0.44,~0,~0)$ for $T_{N2}<T<T_{N1}$, and a commensurate AFM with $\mathbf{q}=(0.5,~0,~0)$ for $T<T_{N2}$\cite{ShanWu2019}. In SDW order, Ce moments might have a small $z$ component that is not shown here. The blue arrows indicate the dipolar field at As sites calculated for specific magnetic structures. (b) Simulated magnetization based on CEF theory shows $\textbf{c}$ as the hard axis, and $\mathbf{ab}$ as the easy plane. The open symbols are experimental results on aligned sample, red squares - $\mathbf{B}_0\perp\mathbf{c}$, blue circles - $\mathbf{B}_0\parallel\mathbf{c}$. The inset depicts CEF splitting. (c) Temperature dependence of magnetic susceptibility measured at $B_0$=10 T, with field applied parallel and perpendicular to $\mathbf{c}$, respectively. The upper inset shows $d(T\chi)/dT$ for non-aligned powder measured at 0.1 T (left), and $C/T$ (right); the lower inset shows $d(T\chi_c)/dT$ for aligned powder measured at 10 T. The two magnetic transitions at $T_{N1}$ and $T_{N2}$ are clearly seen. Both $T_{N1}$ and $T_{N2}$ are relatively lower when under a high field of 10 T.}
\label{Fig1}
\end{figure*}

Unlike the cases for $Tm$=Fe or Co, the $3d$ electrons in the Ni counterpart were shown to be nonmagnetic\cite{XuG-LaOMAs,YongkangLuo2011}. This makes CeNiAsO an ideal candidate to explore the heavy-fermion properties of the $4f$ electrons. Our previous studies based on magnetic susceptibility and specific heat measurements revealed that the Ce moments in CeNiAsO undergo two successive AFM transitions at $T_{N1}\approx9.3$ K and $T_{N2}\approx7.3$ K [see also in Fig.~\ref{Fig1}(c)]. The Sommerfeld coefficient was found to be $\sim 203$ mJ/mol$\cdot$K, indicative of substantial $3d$-$4f$  hybridization. Applying physical pressure to CeNiAsO or substituting As with smaller P ions suppresses the AFM orders and leads to a quantum critical point (QCP) at $p_c\approx0.65$ GPa or $x_c\approx0.4$, respectively\cite{luo2014}. An abrupt change in the Hall coefficient was revealed at $p_c$, characteristic of local (or Kondo-breakdown) quantum criticality \cite{SiQ-localQCP,Coleman-QCP2005,Gegenwart2008,Paschen-YbRh2Si2Hall} around which the suppression of order parameter is also accompanied with breakdown of Kondo effect and hence a Fermi-surface reconstruction. Recently, magnetic structures of the two AFM ordered states were reported by Wu \textit{et al} by a combination of neutron scattering and $\mu$SR experiments\cite{ShanWu2019}. They found that for $T<T_{N1}$, a second order phase transition yields an incommensurate SDW order with a wave vector $\mathbf{q}=(0.44,~0,~0)$; while for $T<T_{N2}$, the magnetic structure evolves into a \textit{coplanar} commensurate AFM order with $\mathbf{q}=(0.5,~0,~0)$. The schematic of these magnetic structures is shown in Fig.~\ref{Fig1}(a). Furthermore, their $\mu$SR measurements on CeNiAs$_{1-x}$P$_x$O manifested that the commensurate order  exists only for $x\leq 0.1$, therefore the QCP is probably between the incommensurate SDW and a paramagnetic phase\cite{ShanWu2019}, which is not expected in the general framework of local quantum criticality. Additional \textit{microscopic local} experiments are invited to confirm the magnetic properties of CeNiAsO.

Herein, we employed $^{75}$As nuclear magnetic resonance (NMR) measurements on field-aligned CeNiAsO powders, and both static and dynamic spin susceptibilities are investigated. NMR spectra confirm two kinds of AFM orderings below $T_{N1}=9.0(3)$ K and $T_{N2}=7.0(3)$ K, respectively. Both NMR shift ($K$) and spin-lattice relaxation rate ($1/T_1$) manifest anisotropic magnetic properties. The hyperfine interaction and the influences of the Kondo effect on $K$ and $1/T_1$ are also discussed.

\section{\Rmnum{2}. Experimental details}

Polycrystalline CeNiAsO sample of high purity was synthesized by solid-state reaction as described previously\cite{YongkangLuo2011}. The sample quality was verified by powder X-ray diffraction (XRD, PANalytical) with Cu K$\alpha$ radiation, and no impurity trace could be detected. Specific heat and low-field (0.1 T) magnetic susceptibility measurements confirm $T_{N1}\simeq9.3$ K and $T_{N2}\simeq7.3$ K [Upper inset to Fig.~\ref{Fig1}(c)], consistent with our original reports. $^{75}$As NMR measurements were performed on field-aligned powders. The polycrystalline sample was thoroughly ground into powders in a glove box, and mixed with Stycast 1266 with a weight ratio $\approx 0.2$ \cite{Young-Alignment}. The mixture was placed in a strong magnetic field of 14 T at 300 K in a physical property measurement system (PPMS-14, Quantum Design) and held for 12 h before the Stycast was cured. Because CeNiAsO is an easy-plane ($\mathbf{ab}$-plane) system (see below), the field alignment was carried out on a rotating sample plate that rotates at a slow speed ($\sim 5$ rpm). $^{75}$As ($^{75}\gamma_n$=7.2919 MHz/T, $I$=3/2) NMR spectra were recorded in a stepped frequency sweep spin-echo method at an external field $B_0\sim 10$ T. The measurements were made for both $\mathbf{B}_0\parallel\mathbf{c}$ and $\mathbf{B}_0\perp\mathbf{c}$. The precise value of $B_0$ was detected by the $^{63}$Cu shift in a second, pre-calibrated coil nearby. Spin-lattice relaxation rate ($1/T_1$) was measured in a standard inversion recovery method on the central ($\frac{1}{2}\leftrightarrow-\frac{1}{2}$) transition, and $T_1$ was extracted by fitting the recovery curve to
\begin{equation}
 M(t)=M(\infty)\{1-2f[\frac{1}{10}\exp{(\frac{-t}{T_{1}})}+\frac{9}{10}\exp{(\frac{-6t}{T_{1}}})]\},
 \label{Eq1}
\end{equation}
where $M(\infty)$, $f$ and $T_1$ are fitting parameters. Anisotropic magnetic susceptibility of CeNiAsO was also measured on the aligned sample with $B_0$=10 T in a PPMS equipped with a vibrating sample magnetometer (VSM).

\section{\Rmnum{3}. Results and Discussion}

\subsection{3.1 CEF effect and magnetic anisotropy}

Before presenting the NMR results, it is necessary to discuss briefly the crystalline electric field (CEF) effect in CeNiAsO. The magnetic anisotropy of a Ce-containing compound is usually described by the CEF effect whose Hamiltonian in $C_{4v}$ point group is
\begin{equation}
\hat{H}_{CEF}=B_2^0 \hat{O}_2^0 + B_4^0 \hat{O}_4^0 + B_4^4 \hat{O}_4^4,
\label{Eq2}
\end{equation}
where $B_{l}^{m}$ ($l$=2,4; $m$=0,4) are CEF parameters, and $\hat{O}_l^m$ are Stevens operators\cite{Hutchings-CEF}. According to Wu's inelastic neutron scattering results\cite{ShanWu2019}, $B_2^0\approx26.7$ K, $B_4^0\approx-1$ K, $B_4^4\approx10.4$ K. It is found that the $j$=5/2 multiplet of Ce$^{3+}$ splits into three Kramers doublets, with the first and second excited doublets sitting at $\sim$226 and 790 K above the ground states, as shown in Tab.~\ref{Tab1} and the inset to Fig.~\ref{Fig1}(b). Simulated field dependent magnetizations at 300 K for field parallel and perpendicular to $\mathbf{c}$-axis are provided in the mainframe of Fig.~\ref{Fig1}(b). One clearly sees $M_c$ much smaller than $M_{ab}$, suggesting $\mathbf{ab}$-plane as the magnetic easy-plane. This confirms the validity of the way we aligned the powdered sample. Based on this simulation, the $g$-factor anisotropy for the ground-state doublet is found to be $g_{\perp c}/g_{\parallel c}=3$.

\begin{table}
\caption{\label{Tab1} CEF parameters, energy levels and wave
functions in CeNiAsO ($C_{4v}$) at zero magnetic field. The CEF parameters are from Ref.~\cite{ShanWu2019}.}
\begin{ruledtabular}
\begin{center}
\def\temptablewidth{1.0\columnwidth}
\begin{tabular}{cc}
\multicolumn{2}{l}{$B^{0}_{2}$=26.7 K, $B^{0}_{4}$=$-$1 K, $B^{4}_{4}$=10.4 K }    \\  \hline
$E_i$ (K) & $|\Psi_i\rangle$ \\ \hline
0         & $|\pm1/2\rangle$ \\
226       & $\mp0.6621|\pm5/2\rangle\pm0.7494|\mp3/2\rangle$ \\
790       & $\mp0.7494|\pm5/2\rangle\mp0.6621|\mp3/2\rangle$ \\
\end{tabular}
\end{center}
\end{ruledtabular}
\end{table}

The temperature dependence of magnetic susceptibility measured under an external field $B_0$=10 T is shown in Fig.~\ref{Fig1}(c). More technical detail about the data analysis of magnetic susceptibility in aligned powders can be found in Ref.~\cite{kangqiaochen2019}. Indeed, $\chi_{c}(T)$ is much smaller than $\chi_{ab}(T)$ in the full window $2 \leq T \leq 300$ K. The two antiferromagnetic transitions are visible in the $d(T\chi_c)/dT$ plot as depicted in the inset to Fig.~\ref{Fig1}(c). Note that the transitions defined by such are relatively lower than in our original report\cite{YongkangLuo2011}, because the large field 10 T likely suppresses both AFM orders. These susceptibility results will be compared with the NMR shifts in the following subsections.

\subsection{3.2 $^{75}$As NMR spectra}

The local environment of $^{75}$As in CeNiAsO can be viewed also in Fig.~\ref{Fig1}(a). Each As is surrounded by four nearest-neighbor Ni (nonmagnetic, 2.365 \AA) and four nearest-neighbor Ce (3.230 \AA). Besides, there are two next-nearest-neighbor Ce sitting right above (4.055 \AA) and below (4.058 \AA) As. We will get back to this later on.

\begin{figure}[!ht]
\vspace{-10pt}
\hspace{-0pt}
\includegraphics[width=9cm]{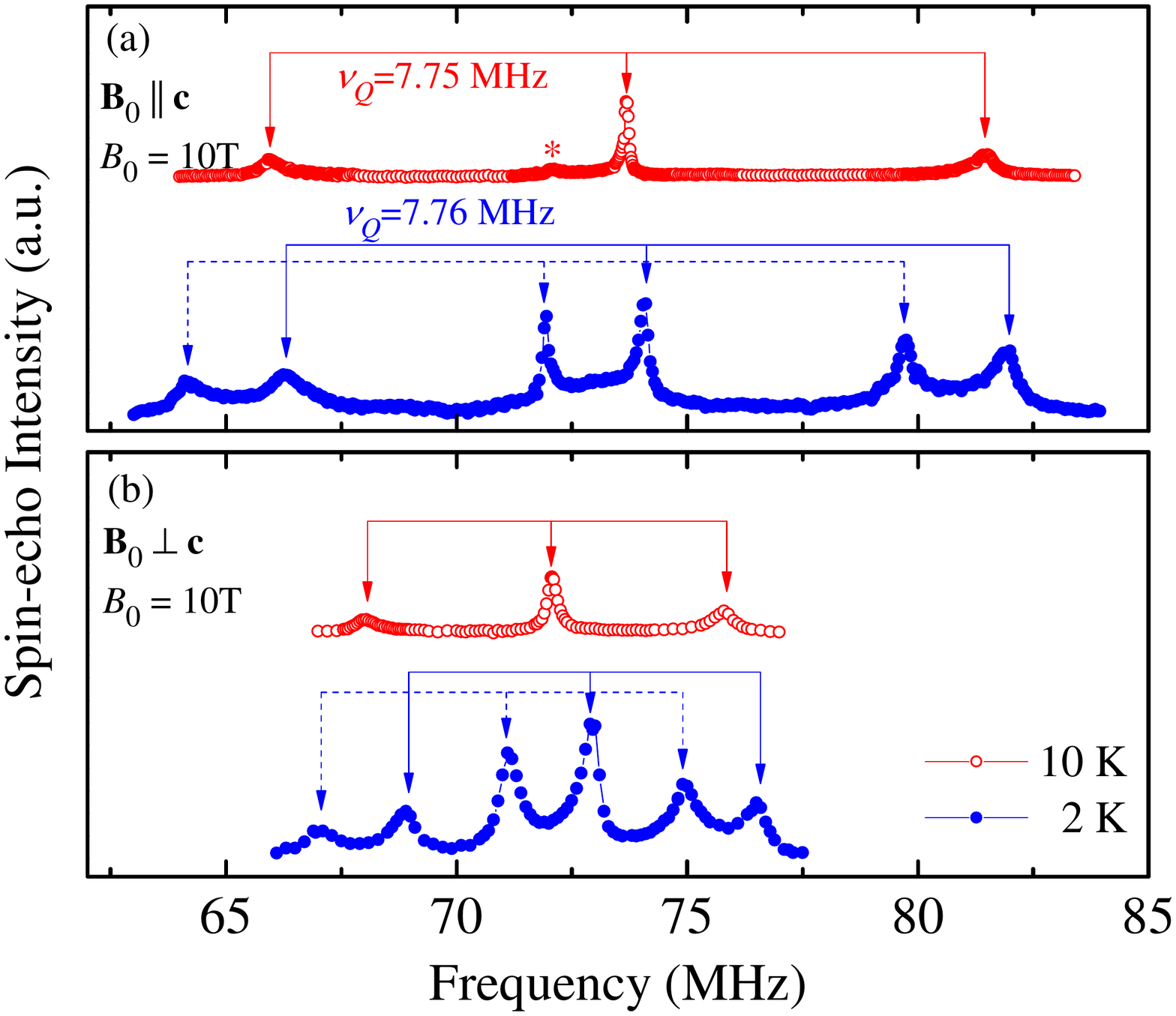}
\vspace{-10pt}
\caption{(a) The $^{75}$As frequency-sweep NMR spectra of CeNiAsO at 10 K (red) and 2 K (blue) with the magnetic field $B_0\approx$ 10 T parallel with $\textbf{c}$. The arrows denote all the central and satellite transitions of the $I$=3/2 $^{75}$As nucleus. The asterisk indicates a small amount of non-aligned powder. (b) ibid, but for field perpendicular to $\textbf{c}$. }
\label{Fig2}
\end{figure}

The full $^{75}$As NMR spectra of aligned CeNiAsO under $B_0 \simeq 10$ T are shown in Fig.~\ref{Fig2}, measured at 10 K (red) and 2 K (blue), respectively. For $\mathbf{B}_0\parallel\mathbf{c}$ and $T=10$ K, three equally separated peaks are resolved at 73.68, 65.96 and 81.46 MHz, corresponding to the central ($\frac{1}{2}\leftrightarrow-\frac{1}{2}$) and satellite ($\pm\frac{3}{2}\leftrightarrow\pm\frac{1}{2}$) transitions of the $I$=3/2 $^{75}$As nucleus. The nuclear quadrupolar splitting frequency is extracted $\nu_Q=$7.75 MHz at 10 K, and remains essentially unchanged down to 2 K, deep inside the AFM ordered phase, manifesting that the principle axis of the electric field gradient (EFG) does not change with AFM transitions. The value of $\nu_Q$ is $\sim 22\%$ smaller than that in CeFeAsO (9.9 MHz)\cite{Ghoshray2009}, suggestive of a drastic change in the charge density distribution around As and hence the character of Fe/Ni-As bonds in these two compounds, which was confirmed previously by band structure calculations\cite{XuG-LaOMAs}. Another salient feature of the spectrum at 2 K is that each peak splits into two peaks, and the splittings are the same among central and satellite transitions, consistent with an AFM ground state at low temperature. We should point out that a small peak is seen at 72.08 MHz [denoted by a asterisk in Fig.~\ref{Fig2}(a)], which should be attributed to a small amount of non-aligned powders. This is demonstrated by the central peak for $\mathbf{B}_0\perp\mathbf{c}$ observed exactly at the same frequency [Fig.~\ref{Fig2}(b)].

\begin{figure}[!ht]
\vspace{-5pt}
\hspace{0pt}
\includegraphics[width=9.8cm]{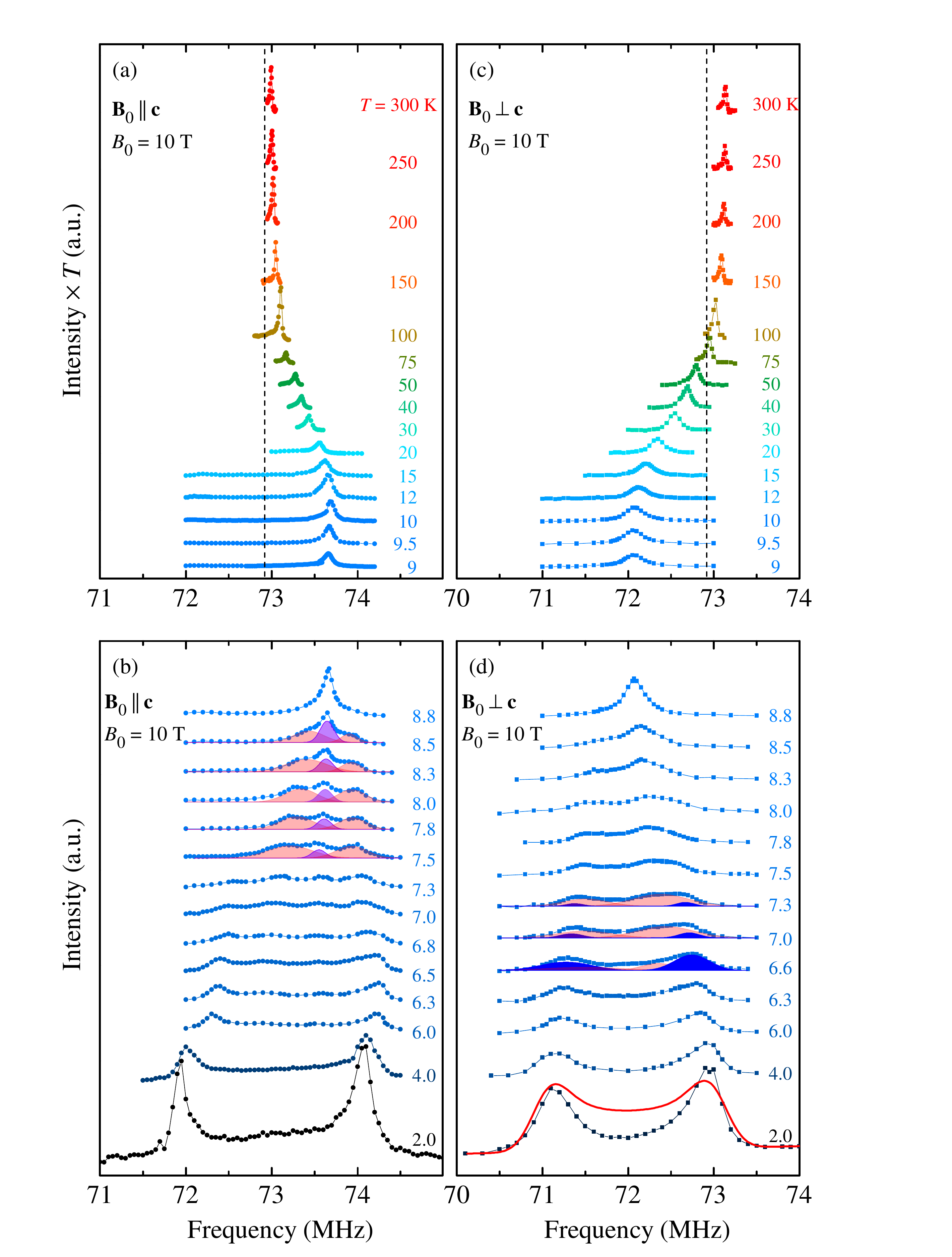}
\vspace{-10pt}
\caption{$^{75}$As frequency-sweep NMR spectra of CeNiAsO at various temperatures under $B_0\approx$ 10 T. (a-b) $\mathbf{B}_0\parallel\mathbf{c}$. (c-d) $\mathbf{B}_0\perp\mathbf{c}$. The dashed lines indicate the Larmor frequency. For clarity, the spectra for $T>T_{N1}$ shown in panels (a) and (c) are amplified by a factor of $T$. The solid red line in (d) is a powder-pattern fitting with Gaussian broadening.}
\label{Fig3}
\end{figure}

Due to the symmetry, we assume $V_{zz}$ is along $\mathbf{c}$-axis, and $V_{xx}=V_{yy}=-\frac{1}{2}V_{zz}$, where $V_{xx}$, $V_{yy}$ and $V_{zz}$ represent the EFG for the three principal axes. Therefore, $\nu_Q'=\nu_Q/2$ is expected for $\mathbf{B}_0\perp\mathbf{c}$. Experimentally, we get $\nu_Q'=3.9(1)$ MHz, as expected [Fig.~\ref{Fig2}(b)]. At 2 K, we also see two sets of peaks due to the AFM ordering. For both field orientations, the central peak splits; it, therefore, is reasonable to infer that the internal field at As sites has both components $\parallel\mathbf{c}$ and $\perp\mathbf{c}$. To show this more directly, we calculated the dipolar fields at each As site exerted by Ce local moments and display them in Fig.~\ref{Fig1}(a) by the blue arrows. In this calculation, the magnetic structure parameters suggested by neutron scattering and $\mu$SR measurements were adopted\cite{ShanWu2019}. At 2 K, the calculated dipolar field is ($-$0.0146, 0, 0.0252) T. Since in our sample the powders are randomly oriented in the $\mathbf{ab}$ plane, we expect there is some distribution of internal field, and this may explain why the split peaks for $\mathbf{B}_0\perp\mathbf{c}$ are broader and with longer tail than for $\mathbf{B}_0\parallel\mathbf{c}$. However, we note that the two well separated AFM peaks ($\mathbf{B}_0\perp\mathbf{c}$) can hardly be fit to a powder pattern with Gaussian broadening, cf Fig.~\ref{Fig3}(d). One possibility might be that the in-plane magnetic anisotropy is not strong, so the application of 10 T field can reorient the moments in the AFM order.

In order to get more detail across the magnetic transitions, we now focus on the central peaks at different temperatures in Fig.~\ref{Fig3}. For better clarity, the spectra are vertically offset, and the position of Larmor frequency $\nu_0=$ $^{75}\gamma_n B_0$ is marked by a dashed line. Magnetic anisotropy can be clearly seen by the opposite movement of peaks at lower temperature, cf Fig.~\ref{Fig3}(a) and (c). For $\mathbf{B}_0\perp\mathbf{c}$, at 300 K, the resonance peak is on the right-hand side of $\nu_0$, and it gradually drifts to the left-hand side upon cooling, indicative that a sign change in NMR shift probably occurs. See below.

Just below $T_{N1}$ and with $\mathbf{B}_0\parallel\mathbf{c}$, we see a shoulder appear on each side of the central peak [cf 8.5 K in Fig.~\ref{Fig3}(b)]. As $T$ decreases, the shoulders evolve into two broad peaks, and meanwhile the height of original central peak in the paramagnetic state shrinks. Since NMR is a microscopic local probe, such an evolution probably indicates phase segregation between the emergent AFM phase and the paramagnetic phase. These patterns can be fit to three Gaussian peaks, and one finds that the two AFM peaks are broad, implying distribution of the internal field in this AFM order, which might be understood by an incommensurate SDW order as proposed by neutron scattering and $\mu$SR experiments\cite{ShanWu2019}. Near 7 K, the spectra are very complicated, and are hard to be fitted to multi-peak functions. Below 6.8 K, two new well separate peaks can be resolved, and this should be associated with the lower-temperature commensurate AFM order suggested by Wu \textit{et al}\cite{ShanWu2019}. The window 6.8-7.3 K probably is a region where these two AFM phases coexist. Such a phase coexistence is also observed for $\mathbf{B}_0\perp\mathbf{c}$, in the similar temperature range [Fig.~\ref{Fig3}(d)]. It is interesting that for this field orientation, we do not see a clear trace of phase segregation just below $T_{N1}$, the reason for which is still unknown.

In spite of some unclear features, two conclusions can be drawn from the temperature dependent $^{75}$As spectra. First, two different AFM orders appear at low temperature, an incommensurate one between $T_{N1}$ and $T_{N2}$, and a lower-temperature phase that seems more likely a commensurate type. Second, for both types of AFM orders, we see peak splittings in both $\mathbf{B}_0\parallel\mathbf{c}$ and $\mathbf{B}_0\perp\mathbf{c}$. This suggests that the internal fields at As sites have both in-plane and out-of-plane components. We will revisit this issue later on.

\subsection{3.3 $^{75}$As NMR shifts and internal field}

\begin{figure}[!ht]
\vspace{-0pt}
\hspace{-0pt}
\includegraphics[width=9.0cm]{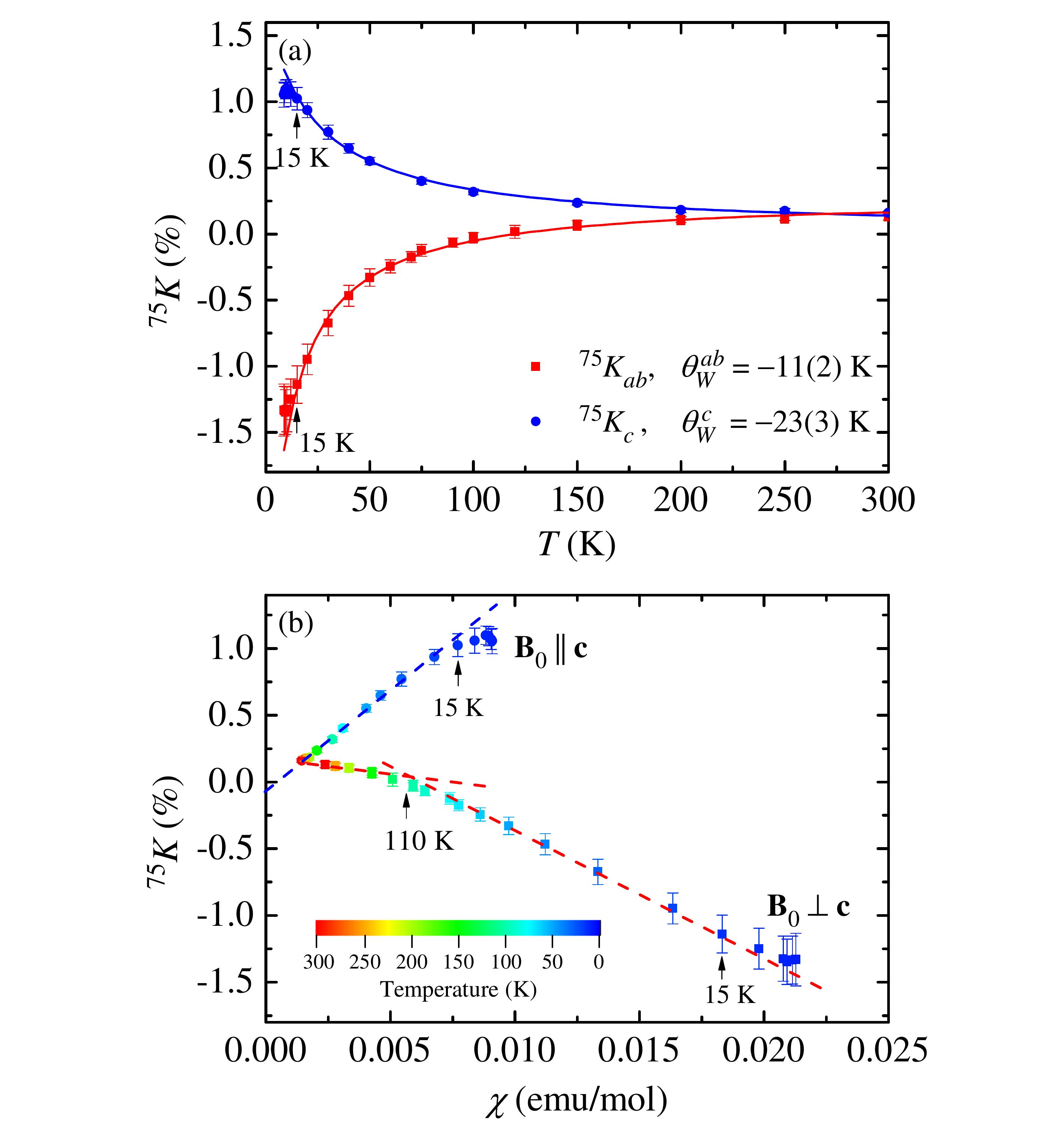}
\vspace{-10pt}
\caption{(a) $^{75}$As NMR shifts in paramagnetic phase as functions of $T$, for field parallel and perpendicular to $\mathbf{c}$. The error bars are determined by the half width at half maximum of NMR peaks. The solid lines are the Curie-Weiss fits to $^{75}K_{ab}(T)$ and $^{75}K_{c}(T)$ in the paramagnetic phase, leading to $\theta_W^{ab}=-11(2)$ K and $\theta_W^{c}=-23(3)$ K, respectively. (b) $^{75}K$ vs. $\chi$ plots with $T$ as an implicit parameter. 
}
\label{Fig4}
\end{figure}

NMR shifts ($K$) of $^{75}$As can be extracted from the central transition lines. Taking account the correction from the second order quadropolar effect\cite{Rybicki2013},
\begin{subequations}
\begin{align}
\nu_{\parallel}&=^{75}\gamma_n B_0(1+K_c),\label{Eq.3a}\\
\nu_{\perp}&=^{75}\gamma_n B_0(1+K_{ab})+\frac{3\nu_Q^2}{16\cdot ^{75}\gamma_n B_0},\label{Eq.3b}
\end{align}
\end{subequations}
and the derived $K_c$ and $K_{ab}$ are displayed in Fig.~\ref{Fig4}(a) as functions of $T$.
Both $K_{c}(T)$ and $K_{ab}(T)$ can be well fit to a Curie-Weiss formula in the $T$ range 20-300 K, with the resultant Weiss temperatures $\theta_W^c=-23(3)$ K and $\theta_W^{ab}=-11(2)$ K, confirming AFM-type exchange couplings between Ce moments. For both field orientations, $^{75}K(T)$ gradually deviates from Curie-Weiss's law below $\sim$ 15 K, probably as a consequence of Kondo effect.

In metals, NMR shifts usually contain multiple contributions,
\begin{equation}
K(T)=K^o+K^s(T),
\label{Eq.4}
\end{equation}
where the temperature independent $K^o$ arises from the orbital shift, and $K^s(T)=A_{hf}\chi(T)$ is the spin (Knight) shift with $A_{hf}$ being the hyperfine coupling constant. A $K-\chi$ plot is given in Fig.~\ref{Fig4}(b). For $\mathbf{B}_0\parallel\mathbf{c}$, this plot gives $A_{hf,c}=0.86(2)$ T/$\mu_B$. Below $\sim$ 15 K, the linear relation between $K(T)$ and $\chi(T)$ gradually collapses. In Kondo lattice compounds, this is usually termed a Knight shift anomaly\cite{Jiang-KAnomaly,Curro_RPP2016}. This provides an estimate to the coherent Kondo scale $T^* \sim 15$ K. We should mention that such a Knight shift anomaly has been observed in a large variety of heavy-fermion materials, and multiple theoretical models have been proposed to attribute the anomalous hyperfine coupling either to a temperature dependence due to Kondo screening\cite{Kim-KAnomaly} or to different occupations of CEF levels\cite{Ohama-CeCu2Si2KAno}. A more recent hypothesis is a phenomenological two-fluid model\cite{Nakatsuji-2Fluid,Yang-PRL2008,Yang-PNAS2012,Shirer-PNAS2012} in which the hyperfine couplings to localized $f$-electron (Fluid 1) spins and itinerant heavy-electron (Fluid 2) spins are of different values. As the weight of the latter starts to increase significantly below $T^*$, $K(T)$ deviates from its linear relation to $\chi(T)$. A similar trend is also seen for $\mathbf{B}_0\perp\mathbf{c}$ near the same temperature, but the feature is relatively weaker. In fact, well above $T^*$, another more prominent shift anomaly is also observed in this field direction at around 110 K. Since this temperature scale is about 50\% of the first CEF splitting energy, it therefore is likely to be caused by CEF depopulation effect\cite{Ohama-CeCu2Si2KAno}. An average value of the hyperfine coupling constant in this field direction is estimated $A_{\text{hf},ab}=-0.42(9)$ T/$\mu_B$.
It should also be noted that in CeNiAsO, the coherent temperature $T^*$ is rather close to the single-ion Kondo scale $T_K$ estimated from our previous entropy analysis\cite{YongkangLuo2011}.

\begin{figure}[!ht]
\vspace{-20pt}
\hspace{-5pt}
\includegraphics[width=8.8cm]{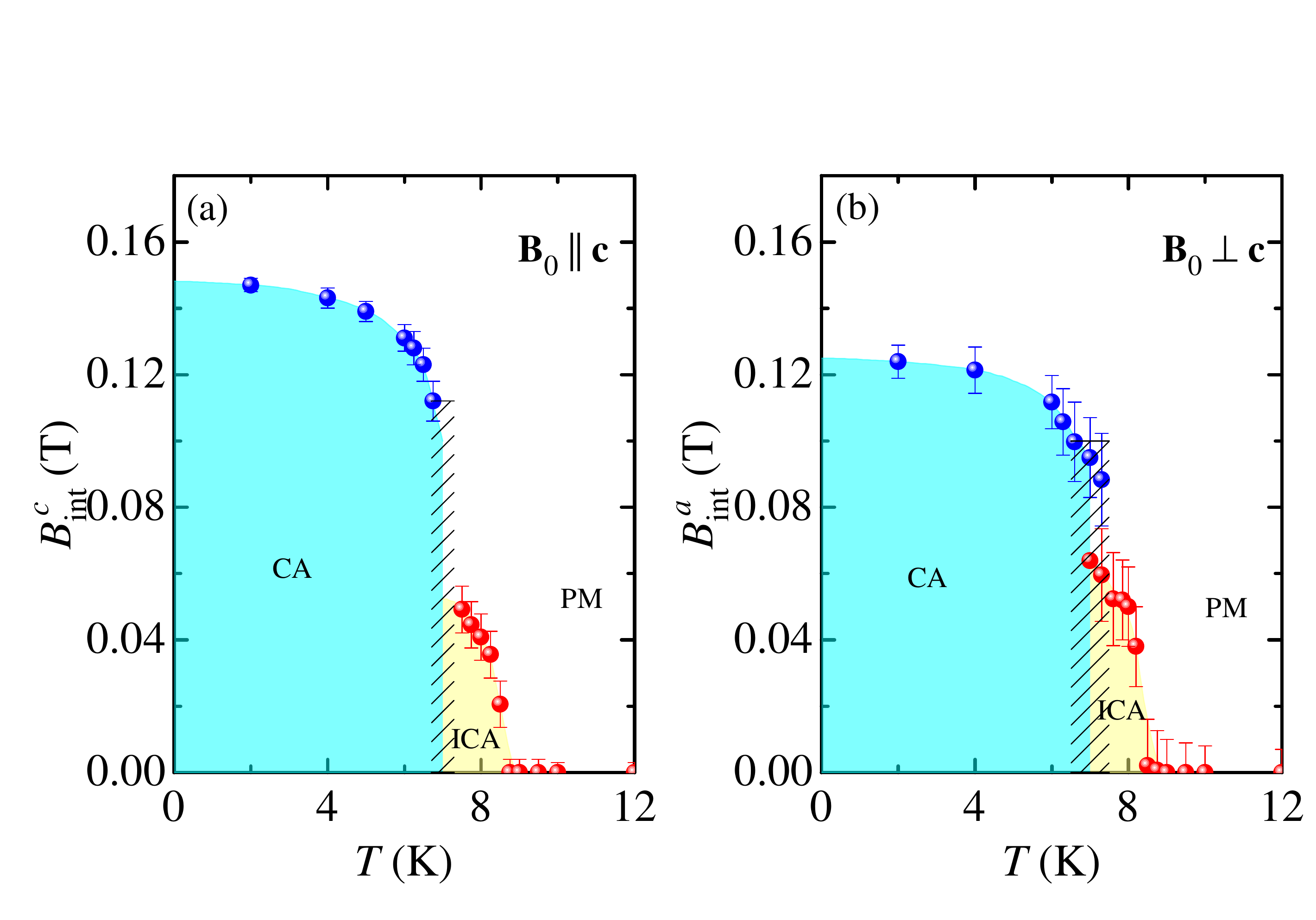}
\vspace{-15pt}
\caption{Internal field at $^{75}$As sites as a function of $T$. (a), $\mathbf{B}_0\parallel\mathbf{c}$. (b), $\mathbf{B}_0\perp\mathbf{c}$. The hatched areas indicate phase coexistence. The abbreviations are: CA = commensurate antiferromagnetic, ICA = incommensurate antiferromagnetic, PM = paramagnetic.}
\label{Fig5}
\end{figure}

Internal fields determined from the AFM-induced splitting (Fig.~\ref{Fig3}) are summarized in Fig.~\ref{Fig5}. Two prominent features are identified. (\rmnum{1}) For both orientations, discontinuities are clearly visible in $B_{int}(T)$ around $T_{N2}$, confirming a modification in magnetic order takes place. (\rmnum{2}) The magnitude of $B_{int}$ is about $\sim$ 6-8 times that of the dipolar field mentioned above (at 2 K). This manifests that the internal field at As sites is dominated by the transferred hyperfine field. Unlike the long-range dipolar interaction, the transferred hyperfine interaction is short-range and relies on orbital hybridization\cite{Kitagawa-Ba122NMR2008}. The ratio $A_{hf,c}/A_{hf,ab}\sim 2$ reminds us that such a hybridization is through the As-$4p_z$ orbital. A tentative try is to consider the four nearest-neighbor Ce sites only, which has the similar symmetry as in the case of LaFeAsO and BaFe$_2$As$_2$ (there each As has four nearest neighbor Fe moments). Following the symmetry analysis in Ref.~\cite{Kitagawa-Ba122NMR2008},
\begin{equation}
\mathbf{B}_{\text{trh}}=\sum_{i=1}^4 \mathbf{A}_i \cdot \mathbf{m}_i,
\label{Eq.5}
\end{equation}
where $\mathbf{B}_{\text{trh}}$ is the transferred hyperfine field, $\mathbf{m}_i$ is the moment at the $i$-th Ce site, and $\mathbf{A}_i$ is the nearest-neighbor hyperfine coupling tensor that takes the form
\begin{equation}
    \mathbf{A}_i=\left[
    \begin{aligned}
        \begin{array}{ccc}
        A_{aa} & A_{ab} & A_{ac} \\
        A_{ba} & A_{bb} & A_{bc} \\
        A_{ca} & A_{cb} & A_{cc} \\
        \end{array}
    \end{aligned}
    \right].
    \label{Eq.6}
\end{equation}


For the lower-temperature AFM phase with wave vector $\mathbf{q}=(0.5,~0,~0)$ \cite{ShanWu2019}, the moments $\mathbf{m}_1$=$-\mathbf{m}_2$=$\mathbf{m}_3$=$-\mathbf{m}_4$=$\mathbf{m}$ \cite{Kitagawa-Ba122NMR2008}, and we get $\mathbf{B}_{\text{trh}}=(4A_{ac}m_z,~0,~4A_{ca}m_x)$. It is interesting to note that in order to have a non-vanishing in-plane hyperfine field as suggested by our NMR measurements (cf Fig.~\ref{Fig5}), the $z$ component of Ce moment should be non-zero. This is different from the coplanar AFM order proposed by neutron and $\mu$SR experiments\cite{ShanWu2019}. Potentially, this bifurcation can be fixed by introducing the next-nearest-neighbor hyperfine coupling tensor, $\mathbf{C}_i$ ($i$=1,2). Since the Ce atoms right above and below are of nearly the same distance to As, for simplicity, we treat them as mirror reflection with respect to $\mathbf{ab}$-plane. By adopting the magnetic structure suggested by Ref.~\cite{ShanWu2019}, we get $\mathbf{B}_{trh}=(4A_{ac}m_z+2C_{aa}m_x-2C_{ab}m_y,~2C_{ba}m_x-2C_{bb}m_y,~4A_{ca}m_x)$. In this case, the in-plane hyperfine field is non-vanishing even $m_z$ is zero. However, we find it difficult to determine the full $\mathbf{A}$ and $\mathbf{C}$ tensors only by the present measurements.


The situation for the incommensurate SDW phase at $T_{N2}<T<T_{N1}$ proposed by Wu \textit{et al}\cite{ShanWu2019} is too complicated for this analysis, and is not discussed here.

\subsection{3.4 Spin dynamics}

\begin{figure}[!ht]
\vspace{-0pt}
\hspace{-0pt}
\includegraphics[width=8.8cm]{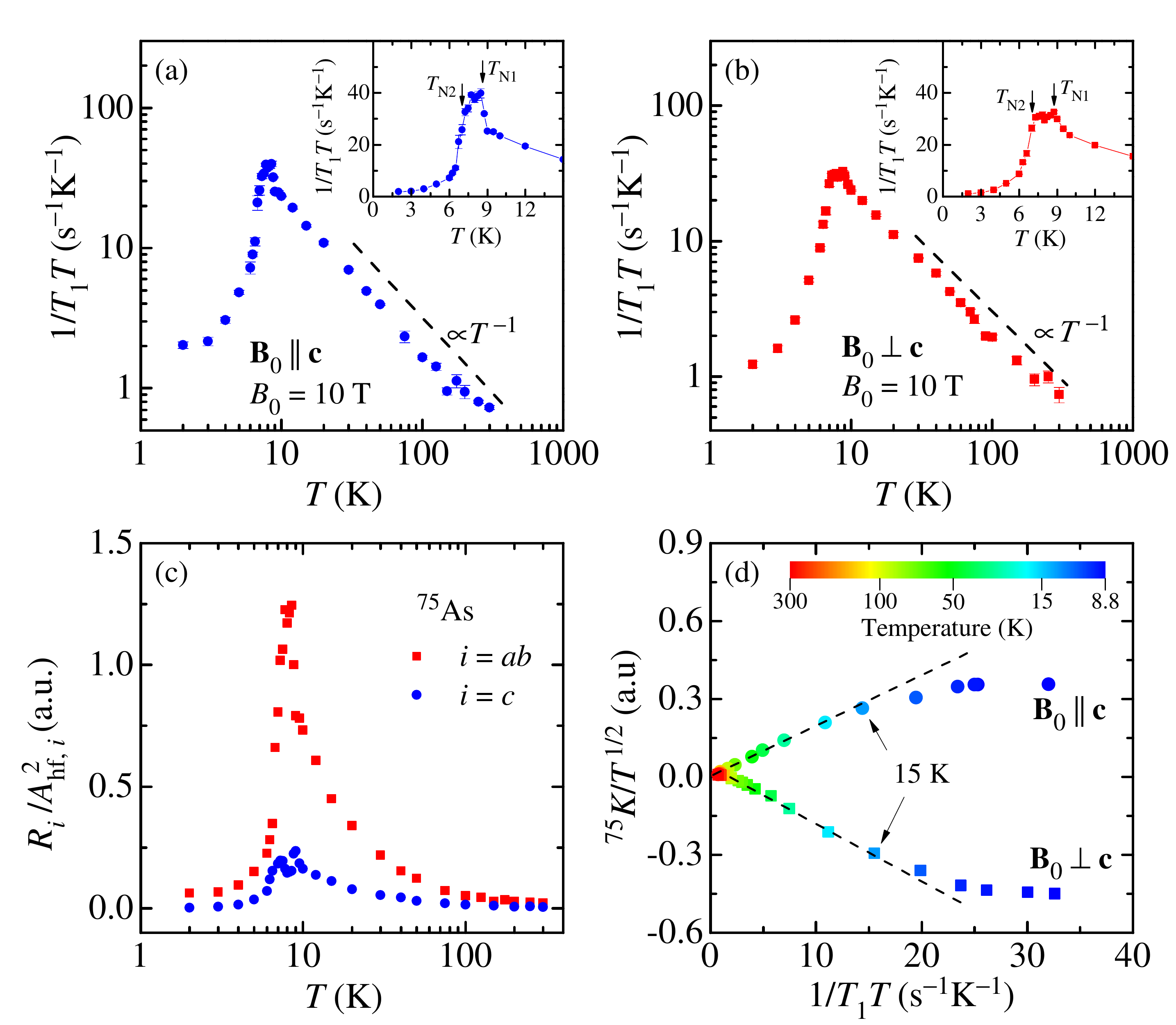}
\vspace{-0pt}
\caption{ (a-b) $T$ dependence of $1/T_1T$ at 10 T, for $\mathbf{B}_0\parallel\mathbf{c}$ and $\mathbf{B}_0\perp\mathbf{c}$, respectively. The insets show $1/T_1T$ near the AFM transitions. (c) Hyperfine-coupling normalized spin fluctuation rate $R_i/A_{\text{hf,}i}^2$ as functions of $T$ for $^{75}$As. (d) Plots of $^{75}K/T^{1/2}$ vs. $1/T_1T$ in the paramagnetic phase for both field orientations.}
\label{Fig6}
\end{figure}

We also investigated the spin dynamics of CeNiAsO by spin-lattice relaxation rate measurements, and the results are summarized in Fig.~\ref{Fig6}. As is well known, for conventional metals (like Cu\cite{Abragam}), the major mechanism to transfer the spin temperature of a nuclear ensemble into equilibrium with the lattice is via spin-flip scattering with conduction electrons\cite{Slichter}; $1/T_1T$, therefore, is a constant proportional to $N^2(E_F)$ [so-called Korringa's relation, where $N(E_F)$ is the density of states at the Fermi level]. In CeNiAsO, however, for both field orientations, $1/T_1$ is essentially constant for $T$ well above $T_{N1}$, or in other words, $1/T_1T \propto T^{-1}$. This is better expressed in the log-log plot shown in Fig.~\ref{Fig6}(a-b). Such a Curie-Weiss-like behavior has been reported in many other Ce compounds like CeIn$_3$\cite{Kohori-CeIn31999},  CePt$_2$In$_7$\cite{Sakai-CePt2In72014}, Ce$_2$CoAl$_7$Ge$_4$\cite{Dioguardi-Ce2CoAl7Ge4NMR}, etc\cite{Bruning-CeFePOFMHF,Rybicki2013}, and can be ascribed to relaxation dominated by spin fluctuations of Ce local moments. Near 9 K, $1/T_1$  peaks for both field orientations, consistent with critical slowing down at the magnetic transitions. The two transition temperatures $T_{N1}$ and $T_{N2}$ are discernible in $1/T_{1}T$, as shown in the inset to Fig.~\ref{Fig6}(a-b). Below $T_{N2}$, $1/T_1$ decreases rapidly, characteristic of long-range magnetic ordering.

In general, the spin-lattice relaxation rate provides a measure of spin fluctuations perpendicular to an applied $\mathbf{B}_0$, as described by\cite{Moriya1974}
\begin{equation}
1/T_1T\propto\gamma_n^2\sum_{\mathbf{q}\perp\mathbf{B}_0}A_{\text{hf}}^2(\mathbf{q},\omega)\frac{\chi''(\mathbf{q},\omega)}{\omega},
\label{Eq.7}
\end{equation}
where $\chi''(\mathbf{q},\omega)$ is the imaginary part of the dynamic susceptibility at wave vector $\mathbf{q}$ and frequency $\omega$. Considering the magnetic anisotropy of the sample, $1/T_1T$ is decomposed for further study:
\begin{subequations}
\begin{align}
(1/T_1T)_{c}&=2R_{ab},\label{Eq.8a}\\
(1/T_1T)_{ab}&=R_{ab}+R_{c},\label{Eq.8b}
\end{align}
\end{subequations}
where the spin fluctuation rate
\begin{equation}
R_{i}\propto A_{\text{hf},i}^2\sum_{\mathbf{q}_{\perp}}\frac{\chi''_i(\mathbf{q},\omega)}{\omega},
\label{Eq.9}
\end{equation}
with $i = ab, c$. Note that here the $\mathbf{q}$ and $\omega$ dependencies of $A_{hf}$ are neglected. The temperature dependence of the hyperfine-coupling normalized spin fluctuation rates $R_{i}/A_{\text{hf},i}^{2}$ are shown in Fig.~\ref{Fig6}(c). $R_{ab}/A_{\text{hf},ab}^{2}$ near $T_{N1}$ is about 5 times as large as $R_{c}/A_{\text{hf},c}^{2}$. This suggests quasi-2D spin fluctuations in CeNiAsO.

Finally, it is necessary to discuss the influence of the Kondo effect on spin dynamics. Previous work on Kondo lattice systems suggested that the $4f$-electron dominated $1/T_1T$ is proportional to the ratio of the static susceptibility $\chi(T)$ and the dynamical relaxation rate $\Gamma(T)$ of the $4f$ electrons\cite{Moriya1974,kuramoto-dynamics2000,maclaughlin-nmr1989,Bruning-CeFePOFMHF}, where $\Gamma\propto T^{1/2}$ for $T>T^*$ \cite{Cox-dynamic1985}. Since $K\propto\chi$ in this temperature range (Fig.~\ref{Fig4}), then
\begin{equation}
(1/T_1T)_{4f}\propto K(T)/T^{1/2}.
\label{Eq.10}
\end{equation}
This relation was found in many Ce-based Kondo compounds including CeFePO\cite{Bruning-CeFePOFMHF} that is isostructural to CeNiAsO. In Fig.~\ref{Fig6}(d), we plot $K/T^{1/2}$ vs. $1/T_1T$ in the paramagnetic phase. For both field directions, Eq.~(\ref{Eq.10}) is validated at high temperature, but is gradually deviated for $T$ below 15 K. This provides additional evidence that a coherent Kondo liquid gradually forms below $T^*\sim15$ K in CeNiAsO.

\subsection{3.5 Discussion and outlook}

Relevant issues remain:

(1) It is still unknown whether the ``phase segregation" observed at $T_{N2}<T<T_{N1}$ is intrinsic or not. First, it is seen only in $\mathbf{B}_0\parallel\mathbf{c}$, but not in $\mathbf{B}_0\perp\mathbf{c}$; second, it disappears when entering the lower-temperature ordered phase; and third, an XRD pattern (not shown here, but equally good as in our original report\cite{YongkangLuo2011}) shows very high sample quality without any noticeable impurity phase. Note that such a phase segregation might not be seen with the resolution of neutron scattering or $\mu$SR. If this phenomenon is intrinsic, considering the heat-capacity results [see in Fig.~\ref{Fig1}(c)] we may infer a weakly first-order transition at $T_{N1}$; it, therefore, is interesting to ask how this order is tuned to a QCP under pressure. However, since the temperature over which the high-$T$ AFM exists is only $\sim 2$ K, comparable to the width for typical phase transitions, and furthermore, within this range, the AFM seems to barely saturate, it, therefore, is premature to make such a conclusion.

(2) A previous study\cite{luo2014} has manifested the realization of QCP in CeNiAsO either by P-doping or pressure. $\mu$SR experiments on CeNiAs$_{1-x}$P$_x$O suggested that the lower-temperature AFM phase is suppressed rapidly by P-doping, and the QCP is between the incommensurate SDW and paramagnetic phases\cite{ShanWu2019}. It is unknown whether the same situation also appears when QCP is accessed by pressure that is a cleaner control parameter than chemical doping. Moreover, an unconventional Kondo-breakdown type QCP was suggested in CeNiAsO\cite{luo2014}; however, the nature of quantum fluctuations (e.g., reflected in scaling laws of Knight shift and spin-lattice relaxation rate\cite{SiQ-localQCP} and whether critical fermionic degrees of freedom should be involved\cite{Prochaska-YbRh2Si2CF,Kandala-QCPCF}, etc) near the QCP remains unclear, and microscopic evidence for a Fermi-surface reconstruction\cite{Kawasaki-Ce115_IrNMR} is also lacking. Hopefully, $^{75}$As NMR under pressure will be helpful to clarify these open questions.

(3) It is also interesting to mention that neither neutron scattering nor $^{75}$As NMR experiment sees any clear trace of structural transition in the AFM ordered states. Of particular interest is that below $T_{N2}$, the wave vector $\mathbf{q}=(0.5,~0,~0)$ is exactly the same as those observed in the parent compounds of iron-based superconductors, whereas in the latter, ubiquitous tetragonal-to-orthorhombic phase transitions were observed\cite{Cruz-La1111Neutron,ZhaoJ-CeFeAsO_F2008,Kitagawa-Ba122NMR2008}. The reason for this difference is unclear. A possibility might be that the magnetoelastic coupling in CeNiAsO is much weaker.

All these require further investigations in the future.

\section{\Rmnum{4}. Conclusions}

By $^{75}$As NMR experiments, we investigated static and dynamic spin susceptibilities of the AFM Kondo lattice material CeNiAsO. The NMR spectra confirm that there are two kinds of AFM ordering that appear below $T_{N1}=9.0(3)$ K and $T_{N2}=7.0(3)$ K, respectively. Quadrupolar splitting frequency $\nu_Q\sim7.75$ MHz remains essentially unchanged across the AFM transitions. A Knight shift anomaly is observed below $T^* \sim 15$ K, which gives a measure of the onset of coherent $c-f$ correlations. Quasi-2D character of spin fluctuations is revealed by the highly anisotropic hyperfine-coupling normalized fluctuation rates. The hyperfine field at As sites transferred from Ce local mements contains both in-plane and out-of-plane components, which highlights the important role played by hybridization to next-nearest-neighbor Ce.

\section{Acknowledgments}

We thank Shan Wu, Stuart E. Brown, Joe D. Thompson, Yifeng Yang, Rui Zhou, and Long Ma for insightful discussions and technical suggestions. This work is supported by the open research fund of Songshan Lake Materials Laboratory (2022SLABFN27), National Key R\&D Program of China (2022YFA1602602), and Guangdong Basic and Applied Basic Research Foundation (2022B1515120020).


%

\end{document}